\documentclass[conference, A4paper]{IEEEtran}
\usepackage{graphicx, amsmath, amsfonts, cite, amssymb, xcolor, subfig, enumerate, mathrsfs,  url, tikz, standalone, mathtools, bbm}
\usetikzlibrary{arrows, patterns}
\newtheorem{lem}{Lemma}
\newtheorem{defi}{Definition}

\newtheorem{prop}{Proposition}

\newtheorem{rem}{Remark}

\newcommand{\be}{\begin{equation}}
\newcommand{\ee}{\end{equation}}
\newcommand{\ben}{\begin{equation*}}
\newcommand{\een}{\end{equation*}}
\newcommand{\mc}{\mathcal}

\newcommand{\abs}[1]{\lvert#1\rvert}

\newcommand{\M}{M}
\newcommand{\Lc}{L_C}
\newcommand{\Lr}{L_R}
\newcommand{\Mc}{M_C}
\newcommand{\Mr}{M_R}

\newcommand{\sfb}{\textsf{b}}
\newcommand{\Wricj}{W_{\textsf{r}(i)\textsf{c}(j)}}
\newcommand{\sfr}{\textsf{r}}
\newcommand{\sfR}{\textsf{R}}
\newcommand{\sfc}{\textsf{c}}
\newcommand{\sfC}{\textsf{C}}
\newcommand{\stdnorm}{\mathcal{N}(0,1)}
\IEEEoverridecommandlockouts

\begin{document}

\title{Spatially Coupled Sparse Regression Codes: \\Design and State Evolution Analysis}
\author{
\IEEEauthorblockN{Kuan Hsieh}
\IEEEauthorblockA{University of Cambridge, UK \\
\tt kh525@cam.ac.uk
}
\and
\IEEEauthorblockN{Cynthia Rush}
\IEEEauthorblockA{Columbia University, USA \\
\tt cynthia.rush@columbia.edu
}
\and
\IEEEauthorblockN{Ramji Venkataramanan}
\IEEEauthorblockA{University of Cambridge, UK \\
\tt ramji.v@eng.cam.ac.uk}
\thanks{This work was supported in part by a Marie Curie Career Integration Grant under grant agreement 631489, and by an EPSRC Doctoral Training Award.}
}
\maketitle

\begin{abstract}
We consider the design and analysis of spatially coupled  sparse regression codes (SC-SPARCs), which were recently introduced by Barbier et al. for  efficient communication over the additive white Gaussian noise channel. SC-SPARCs  can be efficiently decoded using an Approximate Message Passing (AMP) decoder, whose performance in each iteration can be predicted via a set of equations  called state evolution. In this paper, we give an asymptotic characterization of the state evolution equations for SC-SPARCs. For any given  base matrix (that defines the coupling structure of the SC-SPARC) and rate, this characterization can be used to predict whether AMP decoding will succeed in the large system limit. 
We then consider a simple base matrix defined by two parameters $(\omega, \Lambda)$, and show that AMP decoding succeeds in the large system limit for all rates $R < \mc{C}$.  The asymptotic result also  indicates how the parameters of the base matrix affect the decoding progression. Simulation results are presented to evaluate the performance of SC-SPARCs defined with the proposed base matrix. 
\end{abstract}

\section{Introduction} \label{sec:intro}

We consider communication over the memoryless additive white Gaussian noise (AWGN) channel, in which the  output $y$ is generated from input $x$ according to $y = x + w$. The noise  $w$ is Gaussian with zero mean and variance $\sigma^2$, and the input $x$ has an average power constraint $P$. If $x_1, x_2, \ldots ,x_n$ are transmitted over $n$ uses of the channel then
\be\label{eq:average_power_constraint}
\frac{1}{n}\sum_{i=1}^nx_i^2 \leq P.
\ee
The Shannon capacity of this channel is given by $\mc{C} = \frac{1}{2}\ln \left(1+\frac{P}{\sigma^2}\right)$ nats/transmission.

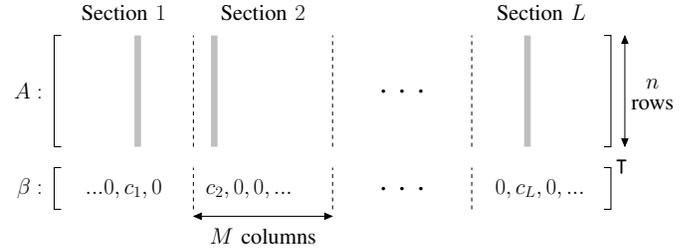
\begin{figure}[t]
    \begin{tikzpicture}[scale=0.375, font=\LARGE]
        \draw[thick] (0.5, 0) -- (0, 0) -- (0, 8) -- (0.5, 8);
        \draw[thick] (39.5, 0) -- (40, 0) -- (40, 8) -- (39.5, 8);
        \draw[line width=5pt, color=gray!50] (6,0) -- (6,8);
        \draw[line width=5pt, color=gray!50] (11.5,0) -- (11.5,8);
        \draw[line width=5pt, color=gray!50] (34,0) -- (34,8);
        \draw[dashed] (10, 0) -- (10, 8);
        \draw[dashed] (20, 0) -- (20, 8);
        \draw[dashed] (30, 0) -- (30, 8);
        \draw (25, 4) node[font=\Huge] {. . .};
        \draw (5, 8.75) node[anchor=south] {Section $1$};
        \draw (15, 8.75) node[anchor=south] {Section $2$};        
        \draw (35, 8.75) node[anchor=south] {Section $L$};
        \draw (-0.5, 4) node[anchor=east] {$A:$};

        \draw[triangle 45-triangle 45] (41, 0) -- (41, 8);
        \draw (43, 4.5) node {$n$};       
        \draw (43, 3) node {rows};

        \draw[thick] (0.5, -4.5) -- (0, -4.5) -- (0, -1.5) -- (0.5, -1.5);
        \draw[thick] (39.5, -4.5) -- (40, -4.5) -- (40, -1.5) -- (39.5, -1.5);
        \draw[dashed] (10, -4.5) -- (10, -1.5);
        \draw[dashed] (20, -4.5) -- (20, -1.5);
        \draw[dashed] (30, -4.5) -- (30, -1.5);
        \draw (-0.5, -3) node[anchor=east] {$\beta:$};
        \draw (25, -3) node[font=\Huge] {. . .};
        \draw (5.0, -3) node {$...0,c_1,0$};
        \draw (14, -3) node {$c_2,0,0,...$};
        \draw (34.9, -3) node {$0,c_L,0,...$};
        \draw (40, -2.25) node[anchor=south west] {$\intercal$};
        \draw[triangle 45-triangle 45] (10, -5) -- (20, -5);
        \draw (15, -5.5) node[anchor=north] {$M$ columns};
    \end{tikzpicture}
\caption{\small $A$ is an $n \times \M L$ design matrix and $\beta$ is an $\M L\times 1$ message vector with one non-zero entry in each of its $L$ sections.  Codewords are of the form $A\beta$. The non-zero values  $c_1, \ldots, c_L$ are fixed a priori.}
\label{fig:sparc_code_construction}
\vspace{-6pt}
\end{figure}

Sparse superposition codes, or sparse regression codes (SPARCs), were  introduced by Joseph and Barron\cite{joseph2012, joseph2014} for efficient communication over the AWGN channel.  These codes have been proven to be reliable at rates approaching $\mc{C}$ with various low complexity iterative decoders \cite{joseph2014,cho2013,rush2017}.  As shown in Fig. \ref{fig:sparc_code_construction}, a SPARC is defined by a design matrix $A$ of dimensions $n\times \M L$, where $n$ is the code length and $\M$, $L$ are integers such that $A$ has $L$ sections with $\M$ columns each. Codewords are generated as linear combinations of $L$ columns of $A$, with one column from each section. Thus a codeword can be represented as $A\beta$, with $\beta$ being an $\M L\times 1$ \emph{message vector} with exactly one non-zero entry in each of its $L$ sections.  The message is indexed by the locations of the non-zero entries in $\beta$. The values of the non-zero entries are fixed a priori. 

Since there are $M$ choices for the location of the non-zero entry in each of the $L$ sections, there are $M^L$ codewords.  To achieve a communication rate of $R$ nats/transmission, we therefore require
\begin{equation}\label{eq:rate_eq}
\M^L = e^{nR} \quad \text{or} \quad nR = L\ln \M.
\end{equation}

In the standard SPARC construction introduced in \cite{joseph2012, joseph2014}, the design matrix $A$ is constructed with i.i.d. standard Gaussian entries. The values of the non-zero coefficients in the message vector $\beta$ then define a \emph{power allocation} across sections.  With an appropriately chosen power allocation (e.g., one that is exponentially decaying across sections),  the feasible decoders proposed in \cite{joseph2014,cho2013,rush2017}  have been shown to be asymptotically capacity-achieving. The choice of power allocation has also been shown to be crucial for obtaining good finite length performance  with the standard SPARC construction  \cite{GreigV17}. 

\emph{Spatially coupled SPARCs}, where the design matrix is composed of blocks with different variances, were recently proposed in \cite{barbier2015,barbier2016,barbier2016itw,BarbierDM17,barbier2017}. In these works, an approximate message passing (AMP)  algorithm was used for decoding, whose performance can be predicted via a recursion known as state evolution. The state evolution recursion was analyzed for a certain class of spatially coupled SPARCs by  Barbier  et al. \cite{barbier2016}, using the potential function method introduced in \cite{yedla2014simple,kumar2014threshold,donoho2013}.   The result in \cite{barbier2016} showed `threshold saturation'  for spatially coupled SPARCs with AMP decoding, i.e., for all rates $R < \mc{C}$,  state evolution predicts vanishing probability of decoding error  in the limit of large code length.
 
As in \cite{barbier2016},   we analyze the AMP decoder for spatially coupled SPARCs via the associated state evolution recursion. However, the analysis in this paper does not use the potential function method; rather, it is based on a simple asymptotic characterization of the state evolution equations.  This characterization gives insight into how the parameters defining the spatial coupling influence the decoding progression. For a given  coupling matrix, the result can be used to determine whether reliable AMP decoding is possible in the large system limit, and the number of  iterations required.  
 
In the rest of the paper, the terminology `large system limit' or `asymptotic limit' refers to $(L,M,n)$ all tending to infinity such that $L \ln M = nR$.

\subsection{Structure of the paper and main contributions}

In Section \ref{sec:sc_AMP}, we review the construction of a spatially coupled SPARC from a base matrix that specifies the variances in the different blocks of the design matrix. In this paper, we use a simple base matrix inspired by protograph-based spatially coupled LDPC constructions \cite{mitchell2015}. This base matrix is defined using two  parameters $(\omega, \Lambda)$,  where $\omega$ is the coupling width and $\Lambda$ is the number of columns in the base matrix.

In Section \ref{sec:AMP} we describe the AMP decoder, and discuss the role of various parameters in the algorithm.
In Section \ref{sec:AMP_perf}, we describe the state evolution equations and  present the main theoretical results. In Lemma \ref{lemma:se_asymp}, we derive an asymptotic expression for the state evolution prediction of the mean-squared error in each iteration.  For any base matrix and rate $R < \mc{C}$, this result can be used to: i)  predict whether the AMP decoder will reliably decode in the large system limit, and ii) compute the number of iterations required by the decoder.    Using this  result, we show  that spatially coupled SPARCs defined via the $(\omega, \Lambda)$ base matrix can reliably decode in the large system limit for all rates $R < \mc{C}$. Furthermore, the result in Proposition \ref{prop:decoding_LSL} also specifies the minimum coupling width $\omega$ and  the number of iterations required for successful decoding. 

In Section  \ref{sec:sims}, we present numerical simulation results to evaluate the  empirical performance of spatially coupled SPARCs, and discuss how the choice of  coupling width in the base matrix can affect finite length decoding performance.

We note that the results in this paper do not constitute a complete proof that spatially coupled SPARCs are capacity-achieving. To prove this, one has to further show that the mean-squared error of the AMP estimates converges almost surely to the corresponding asymptotic state evolution prediction. Obtaining such a result by extending the AMP analysis techniques for standard SPARCs \cite{rush2017,rush2017isit} is part of ongoing work.  

\emph{Notation}: We use $[k]$ to denote the set of the first $k$ integers, $[k] \coloneqq \{1,2,\ldots,k\}$. 
We write $X \sim \mc{N}(\mu, \sigma^2)$  to denote a Gaussian random variable $X$ with mean $\mu$ and variance $\sigma^2$. 


\section{Spatially coupled SPARC construction} \label{sec:sc_AMP}

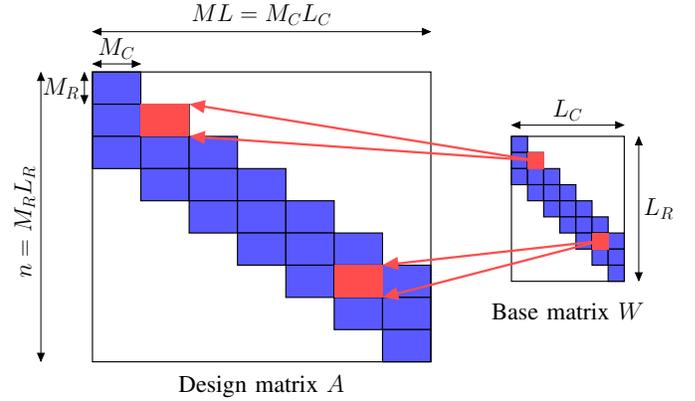
\begin{figure}[t]
\centering
    \begin{tikzpicture}[scale=0.4, font=\LARGE]
	
	\foreach \x in {0,1, 2, 3, 4, 5, 6}{
		\draw[fill=blue!65, thick] (10+3*\x, {-2*\x}) rectangle (13+3*\x, {-2-2*\x});
		\draw[fill=blue!65, thick] (10+3*\x, {-2-2*\x}) rectangle (13+3*\x, {-4-2*\x});
		\draw[fill=blue!65, thick] (10+3*\x, {-4-2*\x}) rectangle (13+3*\x, {-6-2*\x});
	}

        \draw[thick] (10, 0) -- (31, 0) -- (31, -18) -- (10, -18) -- cycle;
	\draw (20.5, -18.5) node[anchor=north] {Design matrix $A$};
	
        \draw[triangle 45-triangle 45] (9.5, 0) -- (9.5, -2);
        \draw (9.5, -1) node[anchor=east] {$\Mr$};
        \draw[triangle 45-triangle 45] (6.8, 0) -- (6.8, -18);
        \draw (5.8, -5.5) node[anchor=east, rotate=90] {$n = \Mr\Lr$};
        \draw[triangle 45-triangle 45] (10, 0.5) -- (13, 0.5);
        \draw (11.5, 0.5) node[anchor=south] {$\Mc$};
        \draw[triangle 45-triangle 45] (10, 2.5) -- (31, 2.5);
        \draw (20.5, 2.6) node[anchor=south] {$\M L = \Mc\Lc$};  	
                     
        \draw[thick] (36, -4) -- (43, -4) -- (43, -13) -- (36, -13) -- cycle; 
        \draw (39.5, -14) node[anchor=north] {Base matrix $W$};
                
        \draw[triangle 45-triangle 45] (43.9, -4) -- (43.9, -13);
        \draw (43.9, -8.5) node[anchor=west] {$\Lr$};
        \draw[triangle 45-triangle 45] (36, -3.3) -- (43, -3.3);
        \draw (39.5, -3.3) node[anchor=south] {$\Lc$};
                
			
	\foreach \x in {0,1, 2, 3, 4, 5, 6}{
		\draw[fill=blue!65, thick] (36+\x, {-4-\x}) rectangle (37+\x, {-5-\x});
		\draw[fill=blue!65, thick] (36+\x, {-5-\x}) rectangle (37+\x, {-6-\x});
		\draw[fill=blue!65, thick] (36+\x, {-6-\x}) rectangle (37+\x, {-7-\x});
	}
	
        	\fill[red!70] (37,-5) rectangle (38,-6);
        	\fill[red!70] (13,-2) rectangle (16,-4);        
	\draw[-triangle 45, red!70, ultra thick] (37.5, -5.5) -- (16, -2);
	\draw[-triangle 45, red!70, ultra thick] (37.5, -5.5) -- (16, -4);
	
	\fill[red!70] (36+5, {-5-5}) rectangle (37+5, {-6-5});
        	\fill[red!70] (25,-12) rectangle (28,-14);
	\draw[-triangle 45, red!70, ultra thick] (41.5, -10.5) -- (28, -12);
	\draw[-triangle 45, red!70, ultra thick] (41.5, -10.5) -- (28, -14);
    \end{tikzpicture}
\caption{\small A spatially coupled design matrix $A$ is divided into blocks of size $\Mr\times \Mc$. There are $\Lr$ and $\Lc$ blocks in each column and row respectively. The independent matrix entries are normally distributed, $A_{ij} \sim \mathcal{N}(0,\frac{1}{L}\Wricj)$, where $W$ is the base matrix. The base matrix shown here is an $(\omega, \Lambda)$ base matrix with parameters $\omega=3$ and $\Lambda=7$. The white parts of $A$ and $W$ correspond to zeros.}
\label{fig:sparc_scmatrix}
\vspace{-6pt}
\end{figure}

Recall from Fig. \ref{fig:sparc_code_construction} that a SPARC  is defined by a design matrix $A$ of dimension  $n\times \M L$, where $n$ is the code length.  In a spatially coupled (SC) SPARC, the matrix $A$ consists of independent zero-mean normally distributed entries whose variances are specified by a \emph{base matrix} $W$ of dimension $\Lr \times \Lc$. The design matrix $A$ is obtained from the base matrix $W$ by replacing each entry $W_{rc}$, for $r\in[\Lr]$, $c\in[\Lc]$, by an $\Mr\times\Mc$ block with i.i.d. entries $\sim \mc{N}(0, W_{rc}/{L} )$.  This is analogous to the ``graph lifting'' procedure in constructing SC-LDPC codes from protographs \cite{mitchell2015}. See Fig. \ref{fig:sparc_scmatrix} for an example, and note that $n= \Lr \Mr$ and $ML = \Lc \Mc$. 

From the construction, the design matrix has independent normal entries
\be\label{eq:construct_Aij}
A_{ij} \sim \mc{N}\left(0,\frac{1}{L} \Wricj \right) \ \forall \ i \in [n], \ j\in[\M L].
\ee
The operators $\sfr(\cdot):[n]\rightarrow[\Lr]$ and $\sfc(\cdot):[\M L]\rightarrow[\Lc]$ in \eqref{eq:construct_Aij} map a particular row or column index to its corresponding \emph{row block} or \emph{column block} index. Conversely, we define operators $\sfR(\cdot)$ and $\sfC(\cdot)$ which map row and column block indices to the \emph{set} of row and column indices they correspond to, i.e.,
\begin{align}
\begin{split}
\sfR(r) &= \{(r-1)\Mr+1,\ldots,r\Mr\} \ \text{for} \ r\in[\Lr], \\
\sfC(c) &= \{(c-1)\Mc+1,\ldots,c\Mc\} \ \text{for} \ c\in[\Lc].
\end{split}
\end{align}
Therefore, $\abs{\sfR(r)} = \Mr$ and $\abs{\sfC(c)} = \Mc$,  for  $r \in [\Lr]$ and $c \in[\Lc]$. We also require  $\Lc$ to divide $L$, resulting in $\frac{L}{\Lc}$ sections per column block.

The non-zero coefficients of $\beta$ (see Fig. \ref{fig:sparc_code_construction})  are all set to $1$, i.e., 
\be c_1 = c_2 = \ldots = c_L = 1.  \ee

For any base matrix $W$, it can be shown that the entries  must satisfy 
\be\label{eq:W_power_contraint}
\frac{1}{\Lr \Lc}\sum_{r=1}^{\Lr}  \sum_{c=1}^{\Lc} W_{rc} = P
\ee
in order to satisfy the average power constraint in  \eqref{eq:average_power_constraint}.

The trivial  base matrix with $\Lr=\Lc=1$  corresponds to a standard (non-SC) SPARC  without power allocation\cite{joseph2012}, while a single-row base matrix $\Lr=1$, $\Lc=L$   is equivalent to standard SPARCs with power allocation \cite{rush2017, joseph2014}. 
In this paper, we will use the following base matrix  inspired by the coupling structure of SC-LDPC codes constructed from protographs \cite{mitchell2015}\footnote{This base matrix construction was also used in \cite{liang2017} for SC-SPARCs. Other base matrix constructions can be found in \cite{barbier2016, barbier2017,donoho2013,krzakala2012}.}. 

\begin{defi}
\label{def:ome_lamb}
An $(\omega , \Lambda)$ base matrix $W$ for SC-SPARCs is  described by two parameters: coupling width $\omega\geq1$ and coupling length $\Lambda\geq 2\omega-1$. The matrix has $\Lr=\Lambda+\omega-1$ rows,  $\Lc=\Lambda$ columns, with each column having $\omega$ identical non-zero entries. For an average power constraint $P$, the  $(r,c)$th entry of the base matrix, for  $r \in [\Lr], c\in[\Lc]$, is given by
\begin{equation}\label{eq:W_rc}
W_{rc} =
\begin{cases}
 	\ P \cdot \frac{\Lambda+\omega-1}{\omega} \quad &\text{if} \ c\leq r \leq c+\omega-1,\\
	\ 0 \quad &\text{otherwise}.
\end{cases}
\end{equation}
\end{defi}
For example, the base matrix in Fig. \ref{fig:sparc_scmatrix} has parameters $\omega=3$ and $\Lambda=7$.

Each non-zero entry in an $(\omega , \Lambda)$ base matrix $W$ corresponds to an $\Mr \times (\M L/\Lc)$ block in the design matrix $A$. Each block can be viewed as a standard (non-SC) SPARC with $\frac{L}{\Lc}$ sections (with $\M$ columns in each section), code length $\Mr$, and rate $R_\text{inner} = \frac{(L/\Lc) \ln \M}{\Mr}$ nats. Using \eqref{eq:rate_eq}, the overall rate of the SC-SPARC is related to $R_\text{inner}$ according to
\be\label{eq:R_Rsparc}
R = \frac{\Lambda}{\Lambda + \omega -1} R_\text{inner}.
\ee
With spatial coupling, $\omega$ is an integer greater than 1, so $R < R_\text{inner}$, which is often referred to as a rate loss. The rate loss depends on the ratio $(\omega-1)/\Lambda$, which becomes negligible when $\Lambda$ is large w.r.t. $\omega$.

\begin{rem}
SC-SPARC constructions generally have a `seed' to jumpstart decoding. In \cite{barbier2016}, a small fraction of sections of $\beta$ are fixed a priori --- this pinning condition is used to analyze the state evolution equations via the potential function method. Analogously, in the construction in \cite{barbier2017}, additional rows are introduced in the design matrix for the blocks corresponding to the first row of the base matrix. In an $(\omega, \Lambda)$ base matrix,  the fact that the number of rows in the base matrix exceeds the number of columns by $(\omega - 1)$ helps decoding start from both ends. The asymptotic state evolution equations derived in Sec. \ref{subsec:asymp_SE} show how AMP decoding progresses in an $(\omega, \Lambda)$ base matrix.
\end{rem}

\section{AMP decoder}\label{sec:AMP}
The decoder aims to recover the message vector $\beta\in\mathbb{R}^{\M L}$ from the channel output sequence $y\in\mathbb{R}^n$, given by
\begin{equation}\label{eq:linear_model}
y = A\beta + w.
\end{equation}

Approximate Message Passing (AMP)  \cite{donoho2009, rangan2010,krzakala2012} refers to a class of iterative algorithms that are Gaussian/quadratic approximations of loopy belief propagation for certain high-dimensional estimation problems (e.g., compressed sensing and low-rank matrix estimation). For decoding SC-SPARCs, we use the following AMP decoder, which is similar to the one used in  \cite{barbier2017}, and can be derived from the Generalized Approximate Message Passing algorithm  \cite{rangan2010} by using the variances specified by the base matrix for the blocks of $A$.  

\begin{figure}[t]
\centering
    \begin{tikzpicture}[scale=0.5, font=\Large]

		\draw[very thick] (0,0) rectangle (6,-1);
		\draw[pattern=north west lines, pattern color=blue, very thick] (0,0) rectangle (1,-1);
		\draw[pattern=dots, pattern color=red, very thick] (1,0) rectangle (2,-1);
		\draw[pattern=crosshatch, pattern color=gray, very thick] (2,0) rectangle (3,-1);
		\draw[pattern=grid, pattern color=green, very thick] (5,0) rectangle (6,-1);
		\draw (4, -0.2) node[anchor=north] {...};
		
		\draw[very thick] (-6,-4) rectangle (12,-5);
		\foreach \x in {0,...,2}{
			\draw[pattern=north west lines, pattern color=blue, very thick] (-6+\x,-4) rectangle (-5+\x,-5);
			\draw[pattern=dots, pattern color=red, very thick] (-3+\x,-4) rectangle (-2+\x,-5);
			\draw[pattern=crosshatch, pattern color=gray, very thick] (\x,-4) rectangle (1+\x,-5);	
			\draw[pattern=grid, pattern color=green, very thick] (12-\x,-4) rectangle (11-\x,-5);
		}
		\draw (6, -4.2) node[anchor=north] {. . . . . .};		
		
		\draw[triangle 45-triangle 45] (0,0.5) -- (6, 0.5);
		\draw (3, 0.7) node[anchor=south] {$\Lr$};
		\draw[triangle 45-triangle 45] (12, -3.5) -- (9, -3.5);
		\draw (10.5, -3.25) node[anchor=south] {$\Mr$};
		\draw[triangle 45-triangle 45] (-6, -5.5) -- (12, -5.5);
		\draw (3, -5.75) node[anchor=north] {$n$};	
		
		\draw[-triangle 45, thick] (3, -1.5) -- (3, -3.5);
		\draw[-triangle 45, thick] (2, -1.5) -- (0.5, -3.5);
		\draw[-triangle 45, thick] (4, -1.5) -- (5.5, -3.5);
		
		\draw (-0.5, -0.32) node[anchor=east] {$\textsf{b}^\intercal:$};
		\draw (-6.5, -4.15) node[anchor=east] {$\widetilde{\textsf{b}}^\intercal:$};		
    \end{tikzpicture}
\caption{$\widetilde{\sfb}^t\in\mathbb{R}^n$ is obtained by repeating $\Mr$ times each entry of $\sfb^t\in\mathbb{R}^{\Lr}$.}
\label{fig:tilde_rep}
\vspace{-5pt}
\end{figure}
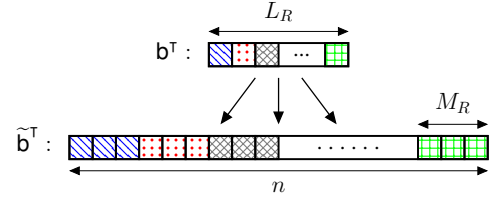

Given the channel output sequence $y$, the AMP decoder generates successive estimates of the message vector, denoted by $\beta^t\in\mathbb{R}^{\M L}$, for $t=0,1,\ldots$. It initialises $\beta^0$ to the all-zero vector, and for $t \geq 0$, iteratively computes
\begin{align}\label{eq:amp_decoder}
\begin{split}
z^t &= y - A\beta^t + \widetilde{\sfb}^t \odot z^{t-1}\\
\beta^{t+1} &= \eta\left(\beta^t + \widetilde{\varsigma}^t\odot\left[ A^\intercal \left(z^t \odot (\widetilde{\varphi}^t)^{-1}\right)  \right], \widetilde{\varsigma}^t \right),
\end{split}
\end{align}
where $\odot$ is the Hadamard (element-wise) product, and $z^{-1}$ is set to the all zero vector. The vector $(\widetilde{\varphi}^t)^{-1}$ denotes the element-wise inverse of $\widetilde{\varphi}^t\in\mathbb{R}^{n}$. The vectors $\widetilde{\sfb}^t\in\mathbb{R}^{n}$ and $\widetilde{\varphi}^t \in\mathbb{R}^{n}$ are obtained by repeating  $\Mr$ times each entry of $\sfb^t\in\mathbb{R}^{\Lr}$ and $\varphi^t\in\mathbb{R}^{\Lr}$  (see Fig. \ref{fig:tilde_rep}). Similarly, $\widetilde{\varsigma}^t\in\mathbb{R}^{\M L}$ is obtained by repeating $\Mc$ times each entry of $\varsigma^t\in\mathbb{R}^{\Lc}$. The vectors $\varsigma^t, \varphi^t, \sfb^t$ in \eqref{eq:amp_decoder} are computed as follows:
\be\label{eq:amp_decoder_tau}
\varsigma^t = \frac{L}{\Mr}\left[W^\intercal (\varphi^t)^{-1}\right]^{-1},
\ee
and for $r\in[\Lr]$,
\be\label{eq:amp_decoder_phi}
\varphi_r^t = \frac{\|z_{\sfR(r)}^t\|_2^2}{\Mr},
\ee
\be\label{eq:amp_decoder_b}
\sfb^t_r = \frac{1}{\Lc} \left[\sum_{c=1}^{\Lc} W_{rc} \left(1 - \frac{\|\beta^t_{\sfC(c)}\|_2^2}{L/\Lc}\right) \right] (\varphi_r^{t-1})^{-1}.
\ee
Finally, let $\text{sec}(\ell)$ denote the set of column indices in the $\ell^{th}$ section, i.e., $\text{sec}(\ell)\coloneqq\{(\ell-1)\M+1,\ldots,\ell \M\}$ for $\ell \in [L]$. The denoising function $\eta: \mathbb{R}^{\M L} \times \mathbb{R}^{\M L} \to \mathbb{R}^{\M L}$ is written as $\eta =(\eta_1, \ldots, \eta_{ML})$, where for $j \in [\M L]$  such that $j\in \text{sec}(\ell)$, 
\be\label{eq:eta_function}
\eta_j(s, \widetilde{\varsigma})
= \frac{e^{s_j/\widetilde{\varsigma}_j}}{\sum_{j'\in \text{sec}(\ell)}e^{s_{j'}/\widetilde{\varsigma}_{j'}}}.
\ee
Notice that $\eta_j(s, \widetilde{\varsigma})$ depends on all the components of $s$ and $\widetilde{\varsigma}$ in the section containing $j$.

When the change in $\beta^t$ (or one of the other parameters) across successive iterations falls below a pre-specified tolerance, or the decoder reaches a maximum allowed iteration number, we take the latest AMP estimate, and set the largest entry in each section to $1$ and the remaining entries to zero. This gives the decoded message vector, denoted  by $\widehat{\beta}$.

\emph{Interpretation of the AMP decoder}: The first argument of the $\eta(\cdot, \cdot)$ in \eqref{eq:amp_decoder}, denoted by $s^t$, can be viewed as a noisy version of $\beta$. In particular, the entry $s^t$ is approximately distributed as 
$\beta + \sqrt{\widetilde{\varsigma}^t}Z$, where $Z$ is a standard normal random vector independent of $\beta$.  Recall that $\beta_{\textsf{C}(c)} \in\mathbb{R}^{\Mc}$ is the part of the message vector corresponding to   column block $c$ of the design matrix. Then, for $c\in[\Lc]$,  the scalar $\varsigma^t_c$  is an estimate of the noise variance in  block $c$ of the effective observation $s^t$, i.e. $\varsigma^t_c \approx \frac{1}{\Mc}\|s_{\textsf{C}(c)}^t-\beta_{\textsf{C}(c)}\|_2^2$.  Under the above distributional assumption, the denoising function $\eta_j$ in \eqref{eq:eta_function} is the  minimum mean squared error (MMSE) estimator  for  $\beta_{j}$, i.e., 
$$ \eta_j(s, \widetilde{\varsigma})=\mathbb{E}\left[\beta_j | s = \beta + \widetilde{\varsigma}Z \right], \qquad \text{ for } j \in [ML],$$
where the expectation is calculated over $\beta$ and  $Z$, with the location of the non-zero entry in each section of $\beta$ being uniformly distributed within the section. 

The vector $z^t\in\mathbb{R}^n$  in \eqref{eq:amp_decoder} is a residual vector, modified with  the `Onsager' term $\widetilde{\sfb}^t \odot z^{t-1}$. This term  arises naturally in the derivation of the AMP algorithm, and is crucial for good decoding performance. For intuition about the role of the Onsager term, see \cite[Sec.~I-C]{bayati2011} and \cite[Sec.~VI]{donoho2013}.
Finally, for $r\in[\Lr]$,  the scalar $\varphi^t_r$ is an estimate of the  variance of the $r$th block of the residual $z^t_{\sfR(r)}$. The residual has approximately zero mean, hence \eqref{eq:amp_decoder_phi} is used to estimate its variance.

The key difference between the AMP decoder in \eqref{eq:amp_decoder} and the one for standard (non-SC) SPARCs   in \cite{rush2017} is that in the latter case, the variance $\varphi^t$ is a scalar that does not depend on  the row index of the base matrix.
\section{Performance of the AMP decoder} \label{sec:AMP_perf}

The performance of a SPARC decoder is  measured by the \emph{section error rate},  defined as
\begin{equation}\label{eq:ser_def}
\mathcal{E}_{\text{sec}} := \frac{1}{L}\sum_{\ell=1}^L \mathbbm{1}\{\widehat{\beta}_{\text{sec}(\ell)} \neq \beta_{\text{sec}(\ell)} \}
\end{equation}
where $\mathbbm{1}$ is the indicator function and $\beta_{\text{sec}(\ell)} $ is the length $\M$ vector corresponding to the $\ell^{th}$ section of the message vector.   If the AMP decoder is run for $T$ steps, then the section error rate can be bounded in terms of the squared error $\| \beta^T - \beta \|^2$. Indeed, since the unique non-zero entry in any section $\ell \in [L]$ equals $1$, we have
\begin{equation}
\label{eq:SER_to_MSE}
\widehat{\beta}_{\text{sec}(\ell)} \neq \beta_{\text{sec}(\ell)} \quad \Rightarrow \quad \|\beta^{T}_{\text{sec}(\ell)} - \beta_{\text{sec}(\ell)} \|_2^2 \geq \frac{1}{4}.
\end{equation}
Recall that $\beta_{\sfC(c)}$ is the part of the message vector corresponding to column block $c$ of the design matrix. There are $\frac{L}{\Lc}$ sections in $\beta_{\sfC(c)}$, with the non-zero entry in each section being equal to  $1$; we denote by $\beta_{\sfC(c)_\ell}$ the $\ell$th of these sections, for $\ell \in [L/\Lc]$. Then, \eqref{eq:SER_to_MSE} implies
\begin{align}
\mathcal{E}_{\text{sec}} & =   \frac{1}{L}\sum_{\ell=1}^L \mathbbm{1}\{\widehat{\beta}_{\text{sec}(\ell)} \neq \beta_{\text{sec}(\ell)} \} \nonumber \\ 
&  =   \frac{1}{L} \sum_{c=1}^{\Lc} \sum_{\ell=1}^{L/\Lc} \mathbbm{1}\left\{\widehat{\beta}_{\sfC(c)_\ell} \neq \beta_{\sfC(c)_\ell} \right\}  \nonumber \\
&  \leq   \frac{4}{L} \sum_{c=1}^{\Lc} \sum_{\ell=1}^{L/\Lc} \|\beta^{T}_{\sfC(c)_\ell} - \beta_{\sfC(c)_\ell}\|_2^2  \nonumber \\
 & =  4 \left[ \frac{1}{\Lc}\sum_{c=1}^{\Lc} \frac{\|\beta^{T}_{\sfC(c)} - \beta_{\sfC(c)} \|_2^2}{L/\Lc}  \right].\label{eq:mse_LB_w_ser}
 \end{align}

We can therefore focus on bounding the \emph{normalized mean square error} (NMSE), which is the bracketed term on the RHS of \eqref{eq:mse_LB_w_ser}.  The normalized mean square error  of the AMP decoder can be predicted using a recursion called state evolution.  For an SC-SPARC defined by base matrix $W$, state evolution (SE) iteratively defines vectors $\phi^t\in\mathbb{R}^{\Lr}$ and $\psi^t\in\mathbb{R}^{\Lc}$ as follows. Initialize $\psi_c^{0} = 1$ for  $c\in[\Lc]$, and for $t=0,1,\ldots$, compute
\begin{align}
\phi_r^t &= \sigma^2 + \frac{1}{\Lc}\sum_{c=1}^{\Lc}W_{rc}\psi_c^t, \qquad r \in [\Lr], \label{eq:se_phi} \\
\psi_c^{t+1} &= 1 - \mathcal{E}(\tau_c^t), \qquad  \qquad c \in [\Lc],  \label{eq:se_psi}
\end{align}
where \be 
\tau_c^t = \frac{R}{\ln{\M}}\left[\frac{1}{\Lr}\sum_{r} \frac{W_{rc}}{\phi_r^t}\right]^{-1}, 
\label{eq:tau_ct_def}
\ee and 
$\mathcal{E}(\tau_c^t)$ is defined as follows.
\be
\label{eq:tau_asymp}
\mathcal{E}(\tau_c^t) 
= \mathbb{E}\left[ \frac{e^{U_1/\sqrt{\tau_c^t}}}{e^{U_1/\sqrt{\tau_c^t}} + e^{- \frac{1}{\tau_c^t}}\sum_{j=2}^{\M} e^{U_j/\sqrt{\tau_c^t}}}\right],
\ee
%
with $U_1,\ldots,U_{\M} \stackrel{\text{i.i.d.}}{\sim} \stdnorm$.    The SE equations in \eqref{eq:se_phi}-\eqref{eq:se_psi} are analogous to those for compressed sensing with spatially coupled measurement matrices  \cite[Eq. (32)-(33)]{donoho2013}, but modified to account for the section-wise structure of the message vector $\beta$.

\begin{figure}[t]
\centering
\includegraphics[width=3.48in]{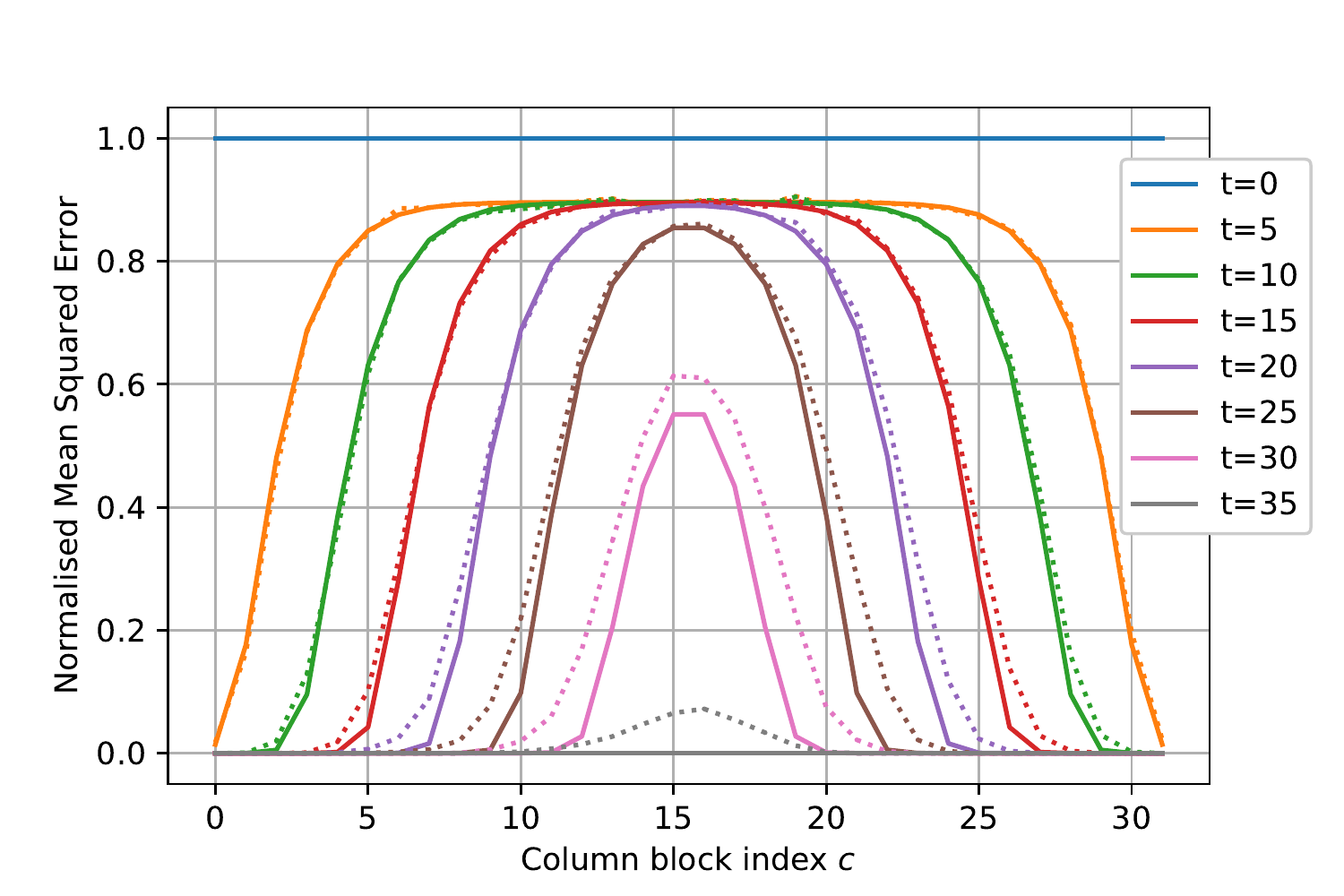}
\caption{ \small NMSE $\frac{\|\beta_{\sfC(c)}^t-\beta_{\sfC(c)}\|_2^2}{L/\Lc}$ vs. column block index $c\in [\Lc]$ for several iteration numbers. The SC-SPARC with an $(\omega,\Lambda)$ base matrix uses parameters: $R=1.5$ bits, $\mc{C}=2$ bits, $\omega=6$, $\Lambda=32$, $M=512$, $L=2048$ and $n=12284$. The solid lines are the SE predictions from \eqref{eq:se_psi}, and the dotted lines are the average NMSE over 100 instances of AMP decoding.}
\label{fig:ser_waveprop}
\vspace{-5pt}
\end{figure}

As illustrated in Fig. \ref{fig:ser_waveprop},  $\psi^t$ closely tracks the NMSE of each block of the message vector, i.e.,  $\psi^t_c \approx \frac{\|\beta_{\sfC(c)}^t-\beta_{\sfC(c)}\|_2^2}{L/\Lc}$ for $c\in[\Lc]$.   Similarly, $\phi^t$ tracks the $\varphi^t$ vector in \eqref{eq:amp_decoder}, which is the block-wise variance of the residual term $z^t$.
We additionally observe from the figure that as AMP iterates, the NMSE  reduction propagates from the ends towards the center blocks.  This decoding propagation phenomenon can be explained using the asymptotic state evolution analysis in the next subsection. 

\subsection{Asymptotic state evolution} \label{subsec:asymp_SE}
Note that $\mathcal{E}(\tau_c^t)$ in \eqref{eq:tau_asymp}  takes a value in $[0,1]$. If $\mathcal{E}(\tau_c^t)=1 $, then $\psi^{t+1}_c =0$, which means that  the sections with indices in $\sfC(c)$ will decode correctly. If we terminate the AMP decoder at iteration $T$, we want  $\psi^{T}_c =0$, for $ c \in [\Lc]$, so that the entire message vector is decoded correctly. The condition under which $\mathcal{E}(\tau_c^t)$ equals 1 in the large system limit is specified by the following lemma.

\begin{lem}\label{lemma:se_asymp}
In the limit as the section size $\M\to\infty$, the expectation $\mathcal{E}(\tau_c^t)$ in \eqref{eq:tau_asymp} converges to either 1 or 0 as follows.
\begin{equation} \label{eq:lim_Etauc}
\lim_{\M\to\infty} \mathcal{E}(\tau_c^t)
	=\begin{cases}
  		1 \quad \text{if} \ \frac{1}{\Lr}\sum_{r=1}^{\Lr} \frac{W_{rc}}{\phi_r^t}>2R\\
		0 \quad \text{if} \ \frac{1}{\Lr}\sum_{r=1}^{\Lr} \frac{W_{rc}}{\phi_r^t}<2R.
	\end{cases}
\end{equation}
This results in the following asymptotic state evolution recursion. Initialise $\bar{\psi}_c^{0} = 1$, for $c\in[\Lc]$, and for $t=0,1,2,\ldots$,
\begin{align}
\bar{\phi}_r^t &= \sigma^2 + \frac{1}{\Lc}\sum_{c=1}^{\Lc}W_{rc}\bar{\psi}_c^t, \qquad r \in [\Lr], \label{eq:se_asmyp_phi} \\
\bar{\psi}_c^{t+1} &= 1 - \mathbbm{1}\left\{ \frac{1}{\Lr}\sum_{r=1}^{\Lr} \frac{W_{rc}}{\bar{\phi}_r^t}>2R \right\},  \qquad c \in [\Lc], \label{eq:se_asmyp_psi}
\end{align}
where $\bar{\phi}, \bar{\psi}$ indicate asymptotic values.
\end{lem}

\begin{rem}
Using the definition of $\tau_c^t$ from \eqref{eq:tau_ct_def}, we can also write \eqref{eq:lim_Etauc} as 
\be
\lim_{\M\to\infty} \mathcal{E}(\tau_c^t)
	=\begin{cases}
  		1 \quad \text{if} \ \tau_c^t \ln \M < \frac{1}{2}\\
		0 \quad \text{if} \ \tau_c^t \ln \M > \frac{1}{2}.
	\end{cases}
\ee
\end{rem}
\begin{IEEEproof}
Recalling the definition of $\tau_c^t$ from \eqref{eq:tau_ct_def}, 
we  write $\frac{1}{\tau_c^t}=\nu_c^t \ln{\M}$, where
\begin{equation}\label{eq:nu_c}
\nu_c^t= \frac{1}{R \Lr}\sum_{r=1}^{\Lr} \frac{W_{rc}}{\phi_r^t}
\end{equation}
is an order 1 quantity because $\frac{1}{\Lr}\sum_{r=1}^{\Lr} W_{rc}=\Theta(1)$. Therefore, 
\begin{equation}
\mathcal{E}(\tau_c^t) = \mathbb{E} \left[\frac{e^{\sqrt{\nu_c^t\ln{\M}}U_1}}{e^{\sqrt{\nu_c^t\ln{\M}}U_1} + {\M}^{-\nu_c^t} \sum_{j=2}^{\M} e^{\sqrt{\nu_c^t\ln{\M}}U_j}}\right],
\end{equation}
which is in the same form as the expectation in Eq. (139) in \cite{rush2017}. Therefore, following the steps in\cite[Appendix~B]{rush2017}, we conclude that
\begin{equation}\label{eq:lemma1_last}
\lim_{\M\to\infty} \mathcal{E}(\tau_c^t)
	=\begin{cases}
  		1 \quad \text{if} \ \nu_c^t>2\\
		0 \quad \text{if} \ \nu_c^t<2.
	\end{cases}
\end{equation}
The proof is completed by substituting the value of  $\nu_c^t$ from \eqref{eq:nu_c} in \eqref{eq:lemma1_last}.
\end{IEEEproof}
%

\begin{rem}
The term $\frac{1}{\Lr}\sum_r \frac{W_{rc}}{\phi^t_r}$ in \eqref{eq:lim_Etauc} represents the average signal to effective noise ratio after iteration $t$ for the column index $c$. If this quantity exceeds the prescribed threshold of $2R$, then the $c^{th}$ block of the message vector, $\beta_{\sfC(c)}$, will be decoded at the next iteration in the large system limit, i.e., $\psi^{t+1}_c=0$.
\end{rem}

The asymptotic SE recursion \eqref{eq:se_asmyp_phi}-\eqref{eq:se_asmyp_psi} is given for a general base matrix $W$. We now apply it to the $(\omega,\Lambda)$ base matrix introduced in Definition \ref{def:ome_lamb}. Recall that an $(\omega,\Lambda)$ base matrix has $\Lr=\Lambda+\omega-1$ rows and $\Lc=\Lambda$ columns, with each column having $\omega$ non-zero entries, all equal to $P \cdot \frac{\Lambda+\omega-1}{\omega}$. 

\begin{lem}\label{lemma:se_asymp_wLbasematrix}
The asymptotic SE recursion \eqref{eq:se_asmyp_phi}-\eqref{eq:se_asmyp_psi} for an $(\omega,\Lambda)$ base matrix $W$ is as follows. Initialise $\bar{\psi}_c^{0} = 1 \ \forall \ c\in[\Lambda ]$, and for $t=0,1,2,\ldots$,
\begin{align}
\bar{\phi}_r^t &= \sigma^2 \left(1 + \frac{\kappa\cdot \text{snr}}{\omega} \sum_{c= \underline{c}_r }^{\overline{c}_r} \bar{\psi}_c^t\right), \quad r \in [ \Lambda + \omega -1], \label{eq:se_asmyp_flat_phi} \\
\bar{\psi}_c^{t+1} &= 1 - \mathbbm{1}\left\{ \frac{P}{\omega}\sum_{r=c}^{c+\omega-1} \frac{1}{\bar{\phi}_r^t}>2R \right\}, \quad c \in [ \Lambda ],  \label{eq:se_asmyp_flat_psi}
\end{align}
where $\kappa = \frac{\Lambda+\omega-1}{\Lambda}$, $\text{snr}=\frac{P}{\sigma^2}$, and
\be\label{eq:c_r}
(\underline{c}_r,\, \overline{c}_r)
	=\begin{cases}
  		(1,\, r) \ &\text{if} \ 1\leq r\leq\omega\\
		(r-\omega+1,\, r) \ &\text{if} \ \omega \leq r \leq \Lambda\\
	        (r-\omega+1,\, \Lambda) \ &\text{if} \ \Lambda \leq r \leq \Lambda + \omega - 1.
	\end{cases}
\ee
\end{lem}
\begin{IEEEproof}
Substitute the value of $W_{rc}$ from \eqref{eq:W_rc}, and $\Lc=\Lambda$, $\Lr=\Lambda+\omega-1$ in \eqref{eq:se_asmyp_phi}-\eqref{eq:se_asmyp_psi}.
\end{IEEEproof}

Observe that the $\bar{\phi}_r^t$'s and $\bar{\psi}_c^t$'s are symmetric about the middle indices, i.e. $\bar{\phi}_r^t = \bar{\phi}_{\Lr-r+1}^t$ for $r\leq \lfloor \frac{\Lr}{2} \rfloor$ and $\bar{\psi}_c^t = \bar{\psi}_{\Lc-c+1}^t$ for $c\leq \lfloor \frac{\Lc}{2} \rfloor$.

Lemma \ref{lemma:se_asymp_wLbasematrix} gives insight into the decoding progression for a large SC-SPARC defined using an  $(\omega,\Lambda)$ base matrix. On initialization ($t=0$),  the value of $\bar{\phi}_r^0$  for each $r$ depends on the number of non-zero entries in row $r$ of $W$, which is equal to $\overline{c}_r - \underline{c}_r + 1$, with $\overline{c}_r, \underline{c}_r$ given by \eqref{eq:c_r}. Therefore, $\bar{\phi}_r^0$ increases from  $r=1$ until $r=\omega$, is constant for $\omega\leq r\leq \Lambda$, and then starts decreasing again after $r = \Lambda$. As a result,  $\bar{\psi}_c^{1}$ is smallest for $c$ at either end of the base matrix ($c\in\{1,\Lambda\}$) and increases as $c$ moves towards the middle, since the $\sum_{r=c}^{c+\omega-1} (\bar{\phi}_r^0)^{-1}$ term in \eqref{eq:se_asmyp_flat_psi} is largest  for $c\in\{1,\Lambda\}$, followed by $c\in\{2, \Lambda-1\}$, and so on. Therefore, we expect the blocks of the message vector corresponding to column index $c\in\{1,\Lambda\}$ to be decoded most easily, followed by $c\in\{2, \Lambda-1\}$, and so on. Fig. \ref{fig:ser_waveprop} shows that this is indeed the case. 

The decoding propagation phenomenon seen in Fig. \ref{fig:ser_waveprop} can also be explained using Lemma \ref{lemma:se_asymp_wLbasematrix} by tracking the evolution of the $\bar{\phi}_r^t$'s and $\bar{\psi}_c^t$'s. In particular, one finds that if column $c^*$ decodes in iteration $t$, i.e. $\bar{\psi}_{c^*}^t=0$, then columns within a coupling width away, i.e. columns $c\in\{c^*-(\omega-1), \ldots, c^*+ (\omega-1)\}$, will become easier to decode in iteration $(t+1)$.

\subsection{Asymptotic State Evolution analysis} \label{subsec:SE_asymp_analysis}

In the following, with  a slight abuse of terminology, we will use the phrase ``column $c$ is decoded in iteration $t$''  to mean $\bar{\psi}_c^t=0$.
\begin{prop}
\label{prop:decoding_LSL}
Consider a SC-SPARC constructed using an $(\omega,\Lambda)$ base matrix with rate $R< \frac{1}{2\kappa}\ln(1+\kappa\cdot\text{snr})$, where $\kappa = \frac{\Lambda+\omega-1}{\Lambda}$. (Note that $\frac{1}{2\kappa}\ln(1+\kappa\cdot\text{snr}) \in [ {\mc{C}}/{\kappa},   \mc{C}]$.) Then, according to the asymptotic state evolution equations in Lemma \ref{lemma:se_asymp_wLbasematrix},  the following statements hold in the large system limit:
\begin{enumerate}
\item The AMP decoder will be able to start decoding if
\be\label{eq:omega_thresh}
\omega > \left( \frac{1}{e^{2R\kappa}-1} - \frac{1}{\kappa\cdot\text{snr}} \right)^{-1}.
\ee
\item If \eqref{eq:omega_thresh} is satisfied, then the  sections in the first and last $c^*$ blocks of the message vector will be decoded in the first iteration (i.e. $\bar{\psi}_c^1=0$ for $c\in\{1,2,\ldots,c^*\}\cup \{\Lambda-c^*+1,\Lambda-c^*+2,\ldots,\Lambda\}$), where $c^*$ is bounded from below as
\begin{align}
 c^* \geq  & \min \Bigg\{  (\omega -1),    \nonumber \\
& \qquad \left\lfloor\omega\cdot\frac{1 + \kappa\cdot\text{snr}}{(\kappa\cdot\text{snr})^2} \cdot 
 \left[ \ln\left(1 + \kappa\cdot\text{snr}\right) - 2R\kappa \right]\right\rfloor  \Bigg\}.\label{eq:c_star}
\end{align}

\item At least $2c^*$ additional columns will decode in each subsequent iteration until the message is fully decoded. Therefore, the AMP decoder will fully decode in at most $\left\lceil \frac{\Lambda}{2c^*} \right\rceil$ iterations.
\end{enumerate}
\end{prop}

\begin{rem}
The proposition implies that for any rate $R < \mc{C}$, AMP decoding is successful in the large system limit, i.e., $\bar{\psi}_c^T=0$ for all $c \in [\Lambda]$. Indeed, consider a rate $R = \mc{C}/\kappa_0$, for any constant $\kappa_0 >1$. Then choose 
$\omega$ to satisfy \eqref{eq:omega_thresh} (with $\kappa$ replaced by $\kappa_0$), and  $\Lambda$ large enough  that $\kappa= \frac{\Lambda + \omega -1}{\Lambda} \leq \kappa_0$. With this choice of $(\omega, \Lambda)$ and rate $R$,  the conditions of the proposition are satisfied, and hence, all the columns decode in the large system limit. 
\end{rem}
\begin{rem}
The proof of the proposition shows that if  $R < \frac{\text{snr}}{ 2(1 + \kappa\cdot\text{snr})}$, then  $\bar{\psi}_c^1=0$, for all $c \in [\Lambda]$, i.e., the entire codeword decodes in the first iteration.
\end{rem}

\begin{IEEEproof}
Since the  $\bar{\phi}_r^t$'s and $\bar{\psi}_c^t$'s in \eqref{eq:se_asmyp_flat_phi} and \eqref{eq:se_asmyp_flat_psi} are symmetric about the middle indices, we will only consider decoding the first half of the columns, $c\in  \{1, \ldots, \lfloor \frac{\Lambda+1}{2} \rfloor \}$, and the same arguments will apply to the second half by symmetry.

In order for column $c$ ($c \leq \lfloor \frac{\Lambda+1}{2} \rfloor$) to decode in iteration 1, i.e. $\bar{\psi}_c^1=0$, we require the argument of the indicator function in \eqref{eq:se_asmyp_flat_psi} to be satisfied for $t=0$, which corresponds to
\begin{align}\label{eq:se_asmyp_flat_t0}
\begin{split}
F_c := \frac{\kappa\cdot\text{snr}}{\omega} \sum_{r=c}^{c+\omega-1} \frac{1}{1 + \frac{\kappa\cdot \text{snr}}{\omega}\cdot(\overline{c}_r- \underline{c}_r+1)} &> 2R\kappa.
\end{split}
\end{align}

1) Since the $F_c$ is largest for column $c=1$, \eqref{eq:se_asmyp_flat_t0} must be satisfied with $c=1$ for \emph{any} column to start decoding. Moreover, using \eqref{eq:c_r}, we find
\begin{align}
F_1 = \frac{\kappa \cdot \text{snr}}{\omega}\sum_{r=1}^{\omega} & \frac{1}{1+ \frac{\kappa \cdot \text{snr}}{\omega}\cdot r}
\stackrel{(\text i)}{>} \int_{\frac{\kappa \cdot \text{snr}}{\omega}}^{\frac{\kappa \cdot \text{snr}}{\omega}(\omega + 1)} \frac{1}{1 + x} \, dx \nonumber \\
&\qquad = \ln\left(1 + \frac{\kappa\cdot\text{snr}}{1+\kappa\cdot\text{snr}\cdot\frac{1}{\omega}}\right), \label{eq:se_asmyp_flat_t0_c1}
\end{align}
where the inequality (i) is obtained by using left Riemann sums on the  decreasing function $\frac{1}{1+x}$. Using \eqref{eq:se_asmyp_flat_t0_c1} in \eqref{eq:se_asmyp_flat_t0}, we conclude that if $\ln\left(1 + \frac{\kappa\cdot\text{snr}}{1+\kappa\cdot\text{snr}/\omega}\right) > 2R\kappa$, then column $c=1$ will decode in the first iteration. 
Rearranging this inequality yields \eqref{eq:omega_thresh}.

2) Given an $(\omega, \Lambda)$ pair that satisfies \eqref{eq:omega_thresh}, we can find a lower bound on the total number of columns that decode in the first iteration. In order to decode column $c$ (and column $\Lambda - c + 1$ by symmetry) in the first iteration, we require \eqref{eq:se_asmyp_flat_t0} to be satisfied. For $c < \omega$, this condition corresponds to
\be\label{eq:se_asmyp_flat_t0_c}
F_{c} = \frac{\kappa\cdot\text{snr}}{\omega} \left[ \left(\sum_{r=c}^{\omega-1} \frac{1}{1 + \frac{\kappa\cdot \text{snr}}{\omega}\cdot r} \right)+ \frac{c}{1 + \kappa\cdot \text{snr}} \right] > 2R\kappa,
\ee
and for columns $c\in\{\omega, \ldots, \Lambda - \omega +1\}$, the condition in \eqref{eq:se_asmyp_flat_t0} reduces to 
\be\label{eq:se_asmyp_flat_t0_omega}
\frac{\text{snr}}{1 + \kappa\cdot\text{snr}} > 2R,
\ee
where \eqref{eq:c_r} was used to find the values of $\underline{c}_r$ and $\overline{c}_r$ . Since $F_c$  defined in \eqref{eq:se_asmyp_flat_t0} is smallest for columns $c\in\{\omega, \ldots, \Lambda - \omega + 1\}$, all columns decode in the first iteration if \eqref{eq:se_asmyp_flat_t0_omega} is satisfied.

For columns $c < \omega$, we can obtain a lower bound for $F_{c}$:
\begin{align}
F_{c}&=\frac{\kappa\cdot\text{snr}}{\omega} \left[ \left(\sum_{r=c}^{\omega-1} \frac{1}{1 + \frac{\kappa\cdot \text{snr}}{\omega}\cdot r} \right)+ \frac{c}{1 + \kappa\cdot \text{snr}} \right] \nonumber \\
&\stackrel{(\text i)}{>} \int_{\frac{\kappa\cdot\text{snr}}{\omega} c }^{ \frac{\kappa\cdot\text{snr} }{\omega} \omega} \frac{1}{1+ x} \, dx + \frac{c}{\omega}\frac{\kappa\cdot\text{snr}}{(1 + \kappa\cdot\text{snr})} \nonumber \\
&= \ln\left(1 + \kappa\cdot\text{snr}\right)  - \ln\left(1+\kappa\cdot\text{snr}\cdot\frac{c}{\omega} \right) + \frac{c}{\omega}\frac{\kappa\cdot\text{snr}}{(1 + \kappa\cdot\text{snr})} \nonumber \\
&\stackrel{(\text{ii})}{>} \ln\left(1 + \kappa\cdot\text{snr}\right)  - \kappa\cdot\text{snr}\cdot\frac{c}{\omega} + \frac{c}{\omega}\frac{\kappa\cdot\text{snr}}{(1 + \kappa\cdot\text{snr})} \nonumber \\
&= \ln\left(1 + \kappa\cdot\text{snr}\right) - \frac{c}{\omega}\frac{(\kappa\cdot\text{snr})^2}{(1 + \kappa\cdot\text{snr})},
\label{eq:se_asmyp_flat_t0_c_LB} %
\end{align}
where (i) is obtained by using left Riemann sums on the decreasing function $\frac{1}{1+x}$, and (ii) from $\ln (x) \leq x-1$. Therefore, if the RHS of \eqref{eq:se_asmyp_flat_t0_c_LB} is greater than $2R\kappa$ then \eqref{eq:se_asmyp_flat_t0_c} is satisfied, and column $c$ will decode in the first iteration. This inequality corresponds to
\begin{equation}\label{eq:c_condition}
c < \omega\cdot\frac{1 + \kappa\cdot\text{snr}}{(\kappa\cdot\text{snr})^2} \cdot \left[ \ln\left(1 + \kappa\cdot\text{snr}\right) - 2R\kappa \right].
\end{equation}
In other words, all columns $c < \omega$ that also satisfy \eqref{eq:c_condition} will decode in the first iteration. Therefore, the number of columns (in the first half) that decode in the first iteration, denoted $c^*$, can be bounded from below by \eqref{eq:c_star}.

3) 
We want to prove that if the first (and last) $c^*$ columns decode in the first iteration, then at least the first (and last) $tc^*$ columns will decode by iteration $t$, for $t \geq 1$. We look at the $c^*<\omega$ case because all columns would have been decoded in the first iteration if $c^* \geq \omega$. We again only consider the first half of the columns (and rows) due to symmetry.

We prove by induction. The $t=1$ case holds by the previous statement that the first $c^*$ columns decode in the first iteration. From \eqref{eq:se_asmyp_flat_t0_c}, this corresponds to the following inequality being satisfied: 
\be\label{eq:se_asmyp_flat_t0_c_star}
\frac{\text{snr}}{\omega} \left[ \left(\sum_{r=c^*}^{\omega-1} \frac{1}{1 + \frac{\kappa\cdot \text{snr}}{\omega}\cdot r} \right)+ \frac{c^*}{1 + \kappa\cdot \text{snr}} \right] > 2R.
\ee

Assume that the statement holds for some $t \geq 1$, i.e. $\bar{\psi}^{t}_c = 0$ for $c\in[tc^*]$. We assume that 
$tc^* < \lfloor \frac{\Lambda +1}{2} \rfloor$, otherwise all the columns will have already been decoded.
Then, from \eqref{eq:se_asmyp_flat_phi}, we obtain
\begin{align}\label{eq:W_flat_phi_tT}
& \bar{\phi}^t_r  \nonumber \\ 
& 	\leq \begin{cases}  
  		\sigma^2, & 1 \leq r\leq tc^*, \\
        		\sigma^2\left( 1 +  \frac{\kappa\cdot\text{snr}}{\omega}(r-tc^*)\right), &  tc^* < r < tc^*+\omega, \\
        		\sigma^2\left( 1 +  \kappa\cdot\text{snr}\right),  &  tc^*+\omega \leq r \leq \lfloor \frac{\Lambda+\omega-1}{2} \rfloor + 1.
	\end{cases}
\end{align}
(We have a $\leq$ sign in \eqref{eq:W_flat_phi_tT} rather than an equality because indices $r$ near  $\frac{\Lambda+\omega-1}{2}$ may have smaller values in the final iterations, due to  columns from the other half and within $\omega$ indices away having already been decoded.)

We now show that the statement holds for $(t+1)$, i.e., $\psi_c^{t+1} =0$ for columns $c\in [(t+1)c^*]$. In order for columns $c\in\{tc^*+1, \ldots, (t+1)c^*\}$ to decode in iteration $(t+1)$, the inequality in the indicator function in \eqref{eq:se_asmyp_flat_psi} must be satisfied when $c=(t+1)c^*$ (the LHS of the inequality is larger for $c\in\{tc^*+1, \ldots, (t+1)c^*-1\}$). This corresponds to
\be\label{eq:se_asmyp_flat_t0_c_starT}
\frac{\text{snr}}{\omega} \left[\left(\sum_{r=(t+1)c^*}^{tc^*+\omega-1} \frac{1}{1 +  \frac{\kappa\cdot\text{snr}}{\omega}(r-tc^*)} \right) + \frac{c^*}{1 +  \kappa\cdot\text{snr}}\right] > 2R,
\ee
which is equivalent to \eqref{eq:se_asmyp_flat_t0_c_star}, noting that $(t+1)c^* < tc^* +\omega$ since $c^*<\omega$. Therefore, \eqref{eq:se_asmyp_flat_t0_c_starT} holds by the condition, and the statement holds for $(t+1)$. Due to symmetry, the same arguments can be applied to the last $tc^*$ and $(t+1)c^*$ columns. Therefore, at least $c^*$ columns from each half will decode in every iteration.
\end{IEEEproof}

\section{Numerical Simulations} \label{sec:sims}

We evaluate the empirical performance of SC-SPARCs constructed from $(\omega, \Lambda)$ base matrices. For the numerical simulations, we use a Hadamard based design matrix instead of a Gaussian one. As demonstrated in \cite{rush2017, barbier2017}, this leads to significant reductions in running time and required memory, with very similar error performance to Gaussian design matrices.

The Hadamard-based design matrix  $A$ is generated as follows.  Let $k = \lceil {\log_2(\max(\Mr+1,\Mc+1))} \rceil$. Each block $A_{\sfR(r),\sfC(c)} \in \mathbb{R}^{\Mr\times\Mc}$,  for $r\in[\Lr], c\in[\Lc]$,   is constructed by choosing $\Mr$ rows uniformly at random\footnote{We do not use the first row and column of the Hadamard matrix because they are all $+1$'s. The other rows and columns have an equal number of $+1's$ and $-1$'s.} from a $2^k\times 2^k$ Hadamard matrix and then scaling each entry up by $\sqrt{\frac{1}{L}W_{rc}}$.  The resulting matrix has entries $A_{ij} = \pm\sqrt{\frac{1}{L}W_{\sfr(i),\sfc(j)}}$. 

\begin{figure}[t]
\centering
\includegraphics[width=3.45in]{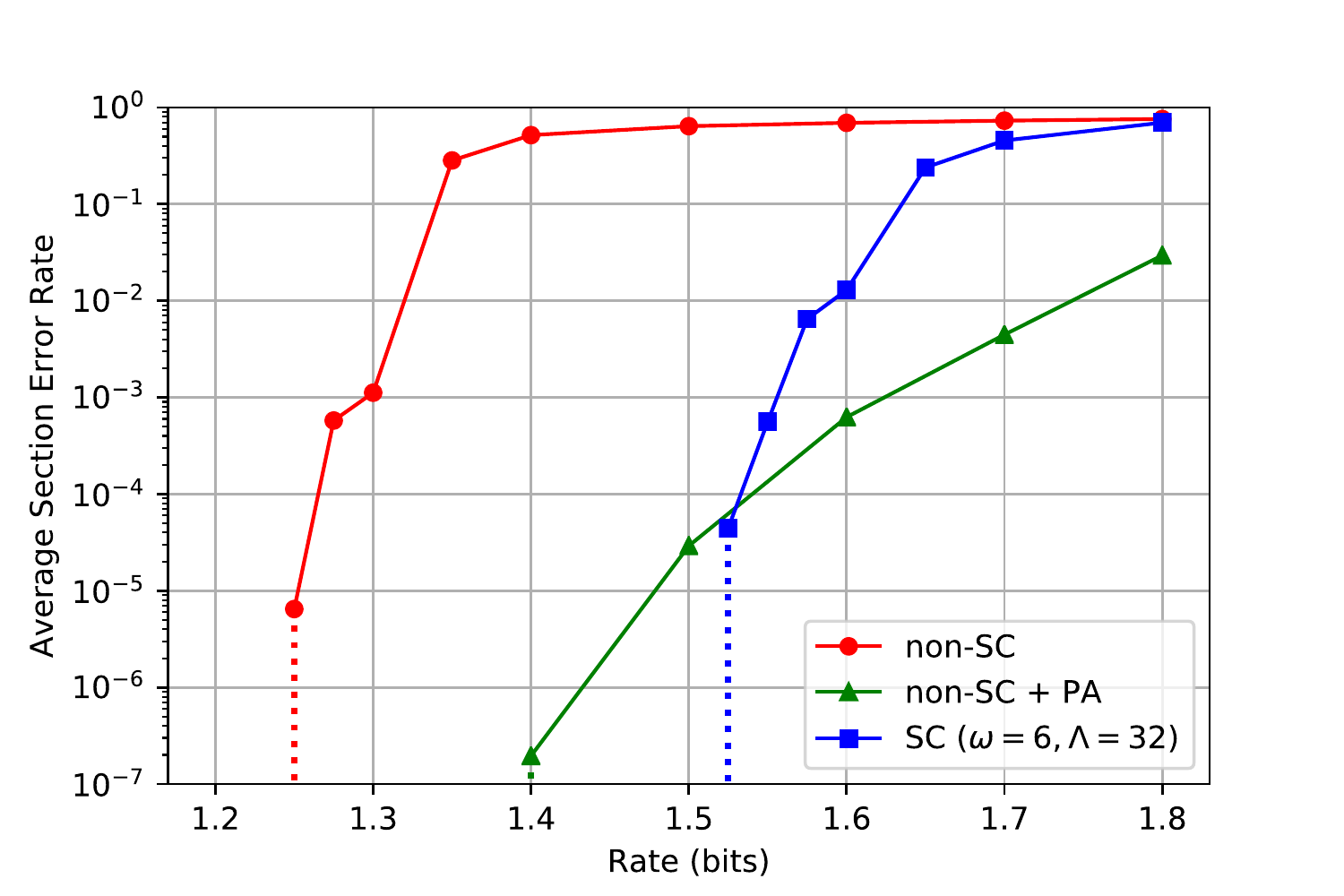}
\caption{ \small Average section error rate (SER) vs. rate at $\text{snr}=15$, $\mc{C}=2$ bits, $\M=512$, $L=1024$, $n\in[5100,7700]$. The SERs are averaged over $10^4$ trials. Plots are shown for non-SC SPARCs with  and without power allocation, and SC-SPARCs with an $(\omega,\Lambda)$ base matrix with $\omega=6,\Lambda=32$. The code length is the same for the three cases. The dotted vertical lines indicate that no section errors were observed over $10^4$ trials at  smaller rates.}
\label{fig:ser4sparcs}
\end{figure}

\begin{figure}[t]
\centering
\includegraphics[width=3.45in]{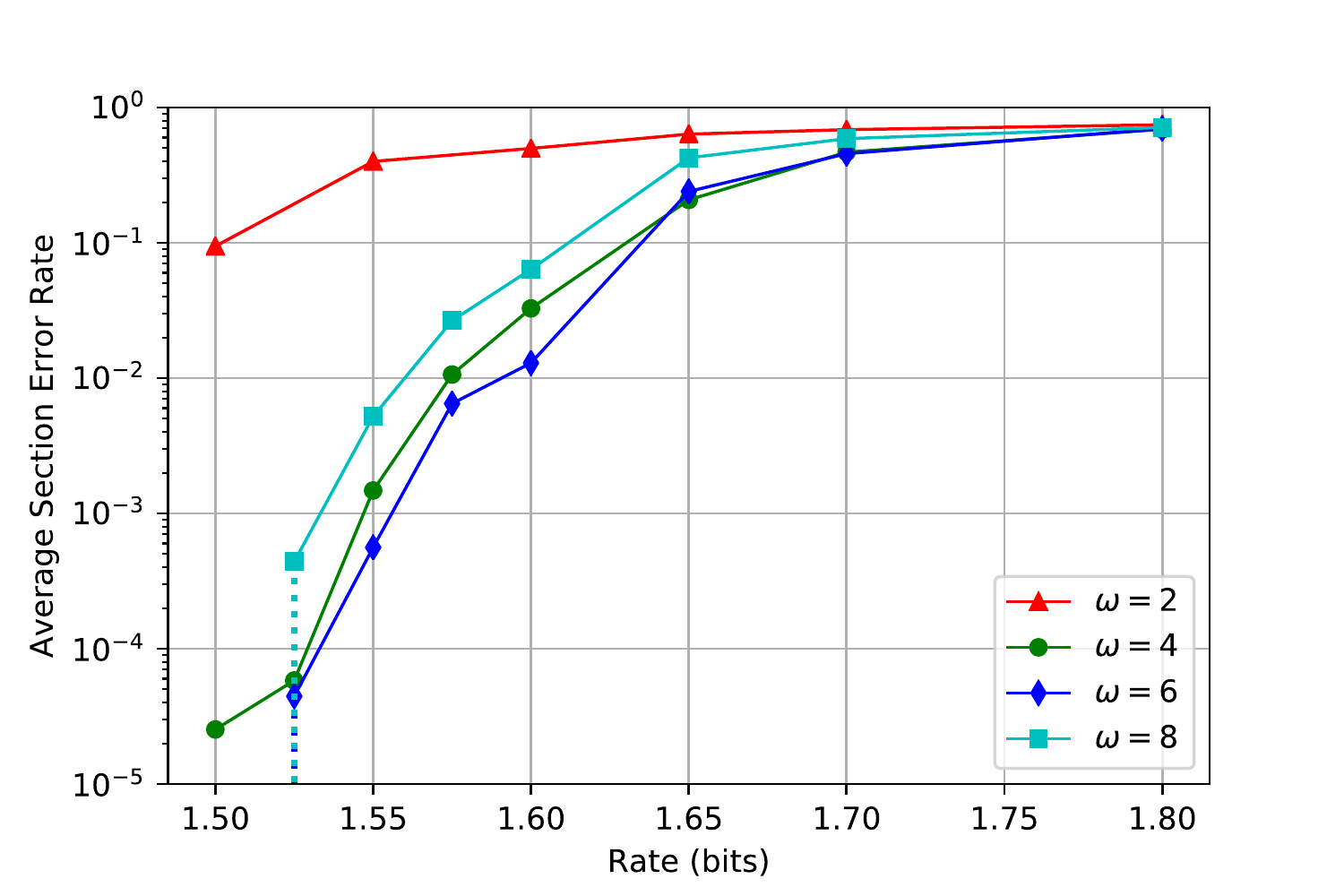}
\caption{\small Average section error rate (SER) vs. rate at $\text{snr}=15$, $\mc{C}=2$ bits, $\M=512$, $L=1024$, $n \in [5100, 6200]$. The SERs are averaged over $10^4$ trials. Plots are shown for SC-SPARCs with an $(\omega,\Lambda)$ base matrix with $\Lambda=32$ and $\omega\in\{2,4,6,8\}$. For a given rate, the code length is the same for different $\omega$ values.  The dotted vertical line indicates that for $\omega=6$ and 8, no section errors were observed over $10^4$ trials at $R=1.5$  bits.}
\label{fig:ser_omega}
\vspace{-5pt}
\end{figure}

Fig. \ref{fig:ser4sparcs} compares the average section error rate (SER) of spatially coupled SPARCs with  standard (non-SC) SPARCs, both with and without power allocation (PA). The code length is the same for all three codes, and the power allocation was designed using the algorithm proposed in \cite{GreigV17}.  AMP decoding is used for all the codes. 
Comparing standard SPARCs without PA and SC-SPARCs, we see that spatial coupling significantly improves the error performance: the rate threshold below which the SER drops steeply to a negligible value is higher for SC-SPARCs. We also observe that at rates close to the channel capacity, standard SPARCs with PA have lower SER than SC-SPARCs. However, as the rate decreases, the drop in SER for standard SPARCs with PA is not as steep as that for SC-SPARCs.

Next, we examine the effect of changing the coupling width $\omega$. Fig. \ref{fig:ser_omega} compares the average SER of SC-SPARCs with $(\omega,\Lambda)$ base matrices with $\Lambda=32$ and varying $\omega$. For a fixed  $\Lambda$, we observe from \eqref{eq:R_Rsparc} that a larger $\omega$ requires a larger inner SPARC rate $R_\text{inner}$ for the same overall SC-SPARC rate $R$. A larger value of $R_\text{inner}$ makes decoding harder; on the other hand increasing the coupling width $\omega$  helps decoding. Thus  for a given rate $R$, there is a trade-off: as illustrated by Fig. \ref{fig:ser_omega}, increasing $\omega$  improves the SER up to a point, but the performance degrades for larger $\omega$.
In general, $\omega$ should be large enough so that coupling can benefit decoding, but not so large that $R_\text{inner}$ is very close to the channel capacity.  For example, for $R=1.6$ bits and $\Lambda=32$, the inner SPARC rate  $R_\text{inner} =1.65, 1.75, 1.85, 1.95$ bits for $\omega=2,4,6,8$, respectively. With  the capacity $\mc{C} =2$ bits,  the figure shows that $\omega=6$  is the best choice for $R=1.6$ bits, with $\omega=8$ being noticeably worse.   This  also indicates that smaller  values $\omega$ would be favored as the rate $R$ gets closer to $\mc{C}$.

\IEEEtriggeratref{5}
\bibliographystyle{ieeetr}
\bibliography{sc_sparcs}

\begin{thebibliography}{10}

\bibitem{joseph2012}
A.~Joseph and A.~R. Barron, ``Least squares superposition codes of moderate
  dictionary size are reliable at rates up to capacity,'' {\em IEEE Trans. Inf.
  Theory}, vol.~58, pp.~2541--2557, May 2012.

\bibitem{joseph2014}
A.~Joseph and A.~R. Barron, ``Fast sparse superposition codes have near
  exponential error probability for {$R<\mathcal{C}$},'' {\em IEEE Trans. Inf.
  Theory}, vol.~60, no.~2, pp.~919--942, 2014.

\bibitem{cho2013}
S.~Cho and A.~R. Barron, ``Approximate iterative {Bayes} optimal estimates for
  high-rate sparse superposition codes,'' in {\em The Sixth Workshop on Inf.
  Theoretic Methods in Sci. and Eng.}, pp.~35--42, 2012.

\bibitem{rush2017}
C.~Rush, A.~Greig, and R.~Venkataramanan, ``Capacity-achieving sparse
  superposition codes via approximate message passing decoding,'' {\em IEEE
  Trans. Inf. Theory}, vol.~63, pp.~1476--1500, March 2017.

\bibitem{GreigV17}
A.~Greig and R.~Venkataramanan, ``Techniques for improving the finite length
  performance of sparse superposition codes,'' {\em IEEE Trans. Commun.},
  vol.~66, pp.~905--917, March 2018.

\bibitem{barbier2015}
J.~Barbier, C.~Sch{\"{u}}lke, and F.~Krzakala, ``{Approximate message-passing
  with spatially coupled structured operators, with applications to compressed
  sensing and sparse superposition codes},'' {\em Journal of Statistical
  Mechanics: Theory and Experiment}, vol.~2015, no.~5, p.~P05013, 2015.

\bibitem{barbier2016}
J.~Barbier, M.~Dia, and N.~Macris, ``Proof of threshold saturation for
  spatially coupled sparse superposition codes,'' in {\em Proc. IEEE Int. Symp.
  Inf. Theory}, 2016.

\bibitem{barbier2016itw}
J.~Barbier, M.~Dia, and N.~Macris, ``{Threshold saturation of spatially coupled
  sparse superposition codes for all memoryless channels},'' {\em Proc. IEEE
  Inf. Theory Workshop}, 2016.

\bibitem{BarbierDM17}
J.~{Barbier}, M.~{Dia}, and N.~{Macris}, ``{Universal Sparse Superposition
  Codes with Spatial Coupling and GAMP Decoding},'' {\em \normalfont ArXiv
  e-prints arXiv:1707.04203}, July 2017.

\bibitem{barbier2017}
J.~Barbier and F.~Krzakala, ``Approximate message-passing decoder and capacity
  achieving sparse superposition codes,'' {\em IEEE Trans. Inf. Theory},
  vol.~63, pp.~4894--4927, Aug 2017.

\bibitem{yedla2014simple}
A.~Yedla, Y.-Y. Jian, P.~S. Nguyen, and H.~D. Pfister, ``A simple proof of
  {M}axwell saturation for coupled scalar recursions,'' {\em IEEE Trans. Inf.
  Theory}, vol.~60, no.~11, pp.~6943--6965, 2014.

\bibitem{kumar2014threshold}
S.~Kumar, A.~J. Young, N.~Macris, and H.~D. Pfister, ``Threshold saturation for
  spatially coupled {LDPC} and {LDGM} codes on {BMS} channels,'' {\em IEEE
  Trans. Inf. Theory}, vol.~60, no.~12, pp.~7389--7415, 2014.

\bibitem{donoho2013}
D.~L. Donoho, A.~Javanmard, and A.~Montanari, ``Information-theoretically
  optimal compressed sensing via spatial coupling and approximate message
  passing,'' {\em IEEE Trans. Inf. Theory}, vol.~59, pp.~7434--7464, Nov. 2013.

\bibitem{mitchell2015}
D.~G.~M. Mitchell, M.~Lentmaier, and D.~J. Costello, ``Spatially coupled {LDPC}
  codes constructed from protographs,'' {\em IEEE Trans. Inf. Theory}, vol.~61,
  pp.~4866--4889, Sept 2015.

\bibitem{rush2017isit}
C.~Rush and R.~Venkataramanan, ``The error exponent of sparse regression codes
  with {AMP} decoding,'' in {\em Proc. IEEE Int. Symp. Inf. Theory}, 2017.

\bibitem{liang2017}
S.~Liang, J.~Ma, and L.~Ping, ``Clipping can improve the performance of
  spatially coupled sparse superposition codes,'' {\em IEEE Commun. Letters},
  vol.~21, pp.~2578--2581, Dec. 2017.

\bibitem{krzakala2012}
F.~Krzakala, M.~M{\'{e}}zard, F.~Sausset, Y.~Sun, and L.~Zdeborov{\'{a}},
  ``{Probabilistic reconstruction in compressed sensing: algorithms, phase
  diagrams, and threshold achieving matrices},'' {\em Journal of Statistical
  Mechanics: Theory and Experiment}, vol.~2012, no.~08, p.~P08009, 2012.

\bibitem{donoho2009}
D.~L. Donoho, A.~Maleki, and A.~Montanari, ``Message-passing algorithms for
  compressed sensing,'' {\em Proceedings of the National Academy of Sciences},
  vol.~106, no.~45, pp.~18914--18919, 2009.

\bibitem{rangan2010}
S.~Rangan, ``Generalized approximate message passing for estimation with random
  linear mixing,'' in {\em Proc. IEEE Int. Symp. Inf. Theory}, 2011.

\bibitem{bayati2011}
M.~Bayati and A.~Montanari, ``The dynamics of message passing on dense graphs,
  with applications to compressed sensing,'' {\em IEEE Trans. Inf. Theory},
  vol.~57, pp.~764--785, Feb 2011.

\end{thebibliography}

\end{document}